%% file: main.tex
\documentclass[journal]{IEEEtran}

\usepackage{booktabs}
\usepackage{array}
\PassOptionsToPackage{table,dvipsnames}{xcolor}
\usepackage{xcolor}
\usepackage{cite}
\usepackage{amsmath,amssymb,amsfonts,bm}
\usepackage{tabularx}
\usepackage{graphicx}
\usepackage{textcomp}
\usepackage[caption=false]{subfig}
\usepackage[hidelinks]{hyperref}
\hypersetup{bookmarksdepth=subsection}
\usepackage{multirow}
\usepackage{adjustbox}
\usepackage{acronym}
\usepackage{pict2e}
\usepackage{siunitx}
\usepackage{enumitem}
\usepackage{balance}

\usepackage{todonotes}

\newsavebox{\ORCIDlogo}
\savebox{\ORCIDlogo}{%
\setlength{\unitlength}{\dimexpr 1em/256\relax}%
\begin{picture}(256,256)%
  \color[HTML]{A6CE39}\put(128,128){\circle*{256}}%
  \color{white}%
  \put(78.6,199.2){\circle*{20}}%
  \moveto(70.9,176,9)\lineto(86.3,176,9)\lineto(86.3,69.8)\lineto(70.9,69.8)%
  \closepath\fillpath%
  \moveto(108.9,176.9)\lineto(150.5,176.9)%
  \curveto(190.1,176.9)(207.5,148.6)(207.5 ,123.3)%
  \curveto(207.5,95,8)(186,69.7)(150.7,69.7)%
  \lineto(108.9,69.7)%
  \closepath\fillpath%
  \color[HTML]{A6CE39}%
  \moveto(124.3,83.6)\lineto(148.8,83.6)%
  \curveto(183.7,83.6)(191.7,110.1)(191.7,123.3)%
  \curveto(191.7,144.8)(178,163)(148,163)%
  \lineto(124.3,163)%
  \closepath\fillpath%
\end{picture}%
}
\newcommand\orcidicon[1]{\href{https://orcid.org/#1}{\usebox{\ORCIDlogo}}}

\usepackage{newtxtext,newtxmath}
\usetikzlibrary{arrows.meta,decorations.pathmorphing,calc,shapes.geometric}
\definecolor{aone}{RGB}{214,96,77}
\definecolor{atwo}{RGB}{244,165,130}
\definecolor{ground}{RGB}{120,160,90}
\definecolor{sat}{RGB}{70,110,170}

\definecolor{lightgrey}{gray}{0.92}
\providecommand{\tabY}{$\bullet$}
\providecommand{\tabN}{$\circ$}
\providecommand{\tabNA}{\textcolor{black!45}{n/a}}
\providecommand{\tabY}{$\bullet$}
\providecommand{\tabH}[1]{\textbf{\begin{tabular}[b]{@{}c@{}}#1\end{tabular}}}

\begin{document}

\title{Cloudy Skies: Early Detection of Jamming\\ Against Iridium LEO Satellite-IoT Links}

\author{Ingrid Huso \ {\orcidicon{0000-0002-5134-5792}} ~\IEEEmembership{Member,~IEEE}, Giulia Fliescher,\\ Savio Sciancalepore\ \orcidicon{0000-0003-0974-3639} ~\IEEEmembership{Member,~IEEE}, and Gabriele Oligeri\ \orcidicon{0000-0002-9637-0430} ~\IEEEmembership{Member,~IEEE}
\thanks{I.~Huso and G.~Oligeri are with the College of Science and Engineering, Hamad Bin Khalifa University, Doha, Qatar (e-mail: ihuso@hbku.edu.qa; goligeri@hbku.edu.qa). G.~Fliescher and S.~Sciancalepore are with the Eindhoven University of Technology and Eindhoven Artificial Intelligence Systems Institute (EAISI), Eindhoven, Netherlands (e-mail: g.fliescher@tue.nl; s.sciancalepore@tue.nl).}
\thanks{This work has been partially supported by the NATO Science for Peace and Security Programme under grant SPS G8414.}
\thanks{The dataset produced in this work is publicly available at~\cite{data}.}
}

\input{acronyms}
\acrodef{JSR}{Jammer-to-Signal Ratio}
\acrodef{NF}{Noise Floor}
\acrodef{RMS}{Root Mean Square}
\acrodef{P95}{95th percentile}

\maketitle

\begin{abstract}
Low Earth Orbit (LEO) satellite-IoT systems increasingly provide connectivity to mobile assets in remote areas, which makes their resilience to radio-frequency jamming a growing concern. Timely jamming detection, i.e., before complete link failure, has been studied for terrestrial networks and Global Navigation Satellite Systems (GNSS), but it remains unclear whether these approaches are effective on LEO links, where propagation varies rapidly, the received signal is weak, and jamming should be detected from the messages that are still correctly received, before the link is lost. This paper investigates the extent to which jamming of Iridium LEO satellite-IoT links can be detected from the physical layer of correctly decoded messages. To this aim, we collect and publicly release a real-world dataset of Iridium messages under 47 jamming conditions, spanning two propagation environments, static and mobile receivers, and different jammer distances, powers, and signal types. 
We further use this dataset to investigate which artifacts introduced by jamming remain observable in correctly decoded Iridium messages. To this end, we analyze three one-class autoencoder-based anomaly detectors operating on physical-layer information and compare them against two signal-quality baseline estimators commonly used as jamming indicators in GNSS-based applications.
Our results show that detectability is governed by the rise of the receiver noise floor, through which jammer power, distance, and receiver mobility act. With 1000-samples chunks, i.e., 40~ms of received signal at 25{,}000 samples per second (collected from about four correctly decoded messages), the autoencoders reach an AUC of at least 0.84 whenever a noise jammer raises the noise floor by at least 9~dB, whereas the $C/N_0$ and SNR estimators fail, since received messages, although correctly decoded, retain a high SNR under jamming. Therefore, timely jamming detection for LEO satellite-IoT links like Iridium should rely on physical-layer information beyond the signal-quality metrics adopted in the GNSS domain.
\end{abstract}

\begin{IEEEkeywords}
Jamming detection, Low Earth Orbit (LEO) satellites, Iridium, physical-layer security, anomaly detection.
\end{IEEEkeywords}

\IEEEpeerreviewmaketitle

\section{Introduction}
\label{sec:intro}
\ac{LEO} satellite constellations increasingly provide connectivity to mobile and remote \ac{IoT} assets that operate beyond the reach of terrestrial infrastructure, including maritime vessels, logistics platforms, environmental monitoring systems, and critical infrastructure~\cite{prol2022_access,hui2025_iotj}. The increasingly low cost of launching and operating small satellites has accelerated this trend, and services such as Iridium offer \ac{M2M} services used in transportation, oil and gas, and construction~\cite{Iridium_satellite_iot}.

At the same time, the reliance on \ac{LEO} links exposes these systems to additional threats~\cite{salkield2026_wisec,coppola2025_usenix}.
Among them, jamming is particularly concerning, since it disrupts communications without compromising protocols or cryptographic mechanisms and can be launched with inexpensive radio equipment~\cite{nguyen2023_wisec}. Detecting it only after the link is lost is often insufficient, especially for mobile devices, since by then operators have already lost connection to the \ac{IoT} device. Therefore, in satellite-\ac{IoT} systems, jamming detection must be \emph{timely}, i.e., detect jamming while operating on messages that are still received and decoded, before losing the link.

Timely jamming detection has been studied for terrestrial links, where semi-supervised one-class \ac{AI} detectors looking for anomalies in raw \ac{PHY}-layer \ac{IQ} samples give promising results~\cite{pirayesh2022_comst,sciancalepore2023_iotj}. Whether these findings transfer to a satellite-\ac{IoT} link is an open question, since the received signal is far weaker and, with satellites orbiting at about \SI{780}{\kilo\meter} and \SI{7.5}{\kilo\meter\per\second}, the geometry, Doppler, and fading of a \ac{LEO} link change within a pass of a few minutes.
Instead, satellite jamming detection has been studied mostly for \ac{GNSS}~\cite{spanghero2025_tifs}, where receivers can detect jamming by monitoring the carrier-to-noise density ratio~\cite{wang2018_gnss}, through blind moment-based~\cite{wiesel_cno} or symbol-directed \ac{SNR} estimators~\cite{irfan2025fingerjam}. 
However, they rely on system assumptions that do not obviously hold for a narrowband \ac{LEO} link like Iridium~\cite{schmidt2016csur}, and no prior work has evaluated them on \ac{LEO} links.

\textbf{Contributions.} In this paper, we investigate to what extent jamming of real-world Iridium satellite-\ac{IoT} links can be identified from the \ac{PHY}-layer information of correctly decoded messages, before link disruption. To this end, we collect a dataset of Iridium messages under 47 jamming conditions and compare the signal-quality indicators used to detect \ac{GNSS} jamming, i.e., \ac{SNR} and $C/N_0$, with \ac{AI}-based anomaly detectors working on raw physical-layer information used on terrestrial links. We make the following contributions:
\begin{itemize}
    \item We show that jamming leaves detectable physical-layer distortions not identifiable by traditional signal-quality metrics, and identify the increase of the receiver noise floor as the primary factor governing jamming detectability in Iridium LEO satellite-IoT links. Through real-world experiments spanning 47 jamming conditions across diverse propagation environments, receiver mobility patterns, jammer power levels, distances, and signal types, we find that, with 1000-samples chunks, the investigated autoencoder-based anomaly detectors achieve an AUC of at least 0.84 whenever noise jamming increases the receiver noise floor by at least 9~dB.
    \item We provide evidence why GNSS-inspired indicators fail when jamming detection operates on correctly decoded messages. In particular, SNR- and $C/N_0$-based estimators often achieve AUC values below 0.7 even under strong interference conditions because received messages, although surviving ongoing jamming, frequently retain favorable signal-quality metrics.
    \item We publicly release a real-world dataset of satellite-IoT data collected under 47 distinct jamming conditions. The dataset is intended to support future research on the security of satellite IoT communication systems.
\end{itemize}

\textbf{Roadmap.} Sec.~\ref{sec:background} reviews the background and the related work, Sec.~\ref{sec:sys_advModel} presents the reference scenario and the adversary model, Sec.~\ref{sec:jem_det_tech} introduces the detectors, Sec.~\ref{sec:data_collection} describes the data collection campaign, Sec.~\ref{sec:evaluation} the evaluation settings and Sec.~\ref{sec:exp_results} the results. Finally, Sec.~\ref{sec:conclusion} concludes the paper.

\begin{table*}[!t]
\centering
\caption{Comparison with the literature on jamming detection: terrestrial (top) and satellite (bottom) links ($\bullet$: yes, $\circ$: no, n/a: not specified).}
\label{tab:comparison}
\setlength{\tabcolsep}{4pt}
\renewcommand{\arraystretch}{1.0}
\begin{adjustbox}{width=\textwidth}
\begin{tabular}{l l l l c c c c c}
\toprule
\textbf{Ref.} & \textbf{Input} & \textbf{Jamming Detection Technique(s)} & \textbf{Jamming signal} & \tabH{Early jamming\\Detection} & \tabH{Semi-\\supervised} & \tabH{Satellite\\link} & \tabH{Real-world\\jamming} & \tabH{Mobile\\RX} \\
\midrule
\multirow{2}{*}{\cite{hanegraaf2025_arxiv}} & IQ & CNN & AWGN, deceptive & \tabY & \tabN & \tabN & \tabY & \tabY \\
 & IQ & AE & AWGN, deceptive & \tabY & \tabY & \tabN & \tabY & \tabY \\
\cite{sciancalepore2023_iotj} & IQ & AE & AWGN, deceptive & \tabY & \tabY & \tabN & \tabY & \tabY \\
\cite{kouchaki2024_wiopt} & IQ & LSTM-AE & \tabNA & \tabN & \tabY & \tabN & \tabN & \tabN \\
\cite{varotto2024_iccw} & Spectrogram & CAE & AWGN & \tabN & \tabY & \tabN & \tabY & \tabN \\
\cite{zahra2023_access} & RSSI, PDR & Rule-based & AWGN & \tabN & \tabNA & \tabN & \tabY & \tabN \\
\cite{Bogdanoski} & RSS & Rule-based & AWGN & \tabY & \tabNA & \tabN & \tabN & \tabN \\
\cite{Garnaev} & Time & Game theory & AWGN & \tabY & \tabNA & \tabN & \tabN & \tabN \\
\cite{Venkata} & PDR & SVM & AWGN & \tabY & \tabN & \tabN & \tabN & \tabN \\
\cite{Villain} & Spectral data & AHC & AWGN & \tabY & \tabY & \tabN & \tabY & \tabN \\
\cite{gecgel2021_wiseml} & OFDM frames & CNN, SVM & AWGN & \tabN & \tabN & LEO & \tabN & \tabN \\
\cite{vazquez2021_commag} & IQ & AE & \tabNA & \tabN & \tabY & GEO & \tabN & \tabN \\
\cite{smailes2024_spacecom} & IQ & AE & AWGN & \tabN & \tabY & LEO & \tabY & \tabN \\
\midrule
\rowcolor{lightgrey}
\textbf{This work} & IQ & Raw, IB, LSTM AEs; $C/N_0$, SNR est. & AWGN, deceptive & \tabY & \tabY & LEO & \tabY & \tabY \\
\bottomrule
\end{tabular}
\end{adjustbox}
\end{table*}

\section{Background and Related Work}
\label{sec:background}


\textbf{The Iridium Constellation.} Iridium operates in \ac{LEO} at approximately \SI{780}{\kilo\meter} above Earth. Its 66 satellites are cross-linked with up to four neighbour satellites, providing redundancy and near-global coverage~\cite{Iridium_web}. The user link occupies the [1616--1626.5]~MHz band and uses \ac{QPSK}, a low-order modulation that trades throughput for robustness to noise~\cite{oligeri2023_tifs}, whose symbols are represented as complex \ac{IQ} samples $I+jQ$, with $I$ the in-phase and $Q$ the quadrature component~\cite{theodore2024_book}.
Most messages are encrypted and authenticated, but a few channels, notably the \ac{IRA} channel, are broadcast in clear. Such messages can be received and decoded with a low-cost \ac{SDR}, a general-purpose active antenna, and open-source toolkits such as GNURadio and \emph{gr-Iridium}~\cite{muccc_grIridium_2024}. In line with recent \ac{LEO} security work~\cite{oligeri2023_tifs,smailes2024_spacecom,wigchert2025_comnet}, we use Iridium as our reference constellation, and its machine-to-machine service as our IoT reference application.

\textbf{AI-based Tools.} To detect anomalies in the received data, we use \acp{AE}, i.e., neural networks trained to reconstruct its input. Its encoder maps the input $x$ to a latent vector $z = f_\theta(x) \in \mathbb{R}^d$, and its decoder returns the reconstruction $\hat{x} = g_\phi(z)$, with the parameters chosen to minimize a reconstruction error, typically the \ac{MSE} over the $n$ input features~\cite{michelucci2022_arxiv}. Trained on samples of a single distribution, an \ac{AE} reconstructs them with small errors and samples of a different distribution with larger ones, so that a threshold $\tau$ on the error turns it into a \ac{OCC} anomaly detector. We also use \ac{LSTM} networks~\cite{hochreiter1997}, i.e., \acp{RNN} that carry information across a sequence through a memory cell and input, output, and forget gates, which mitigate the vanishing and exploding gradient problem~\cite{mathworks}. Combined with an \ac{AE}, they yield the \ac{LSTM-AE}, which learns dependencies along a sequence from one class of data only and flags an input as anomalous when its reconstruction error exceeds a threshold.

\textbf{Related Work.} Jamming detectors process either link metrics or \ac{PHY}-layer data. Most recent works use \ac{IQ} samples~\cite{hanegraaf2025_arxiv,kouchaki2024_wiopt,sciancalepore2023_iotj,smailes2024_spacecom,vazquez2021_commag}, while others use spectrograms~\cite{varotto2024_iccw}, \ac{OFDM} WiFi frames~\cite{gecgel2021_wiseml}, or both link and \ac{PHY}-layer metrics~\cite{zahra2023_access}. For the low-order modulations of \ac{LEO} links, such as \ac{BPSK} and \ac{QPSK}, the \ac{CSI} used by some of these approaches becomes unreliable, which makes \ac{IQ} samples the natural input.
Among \ac{AI}-based techniques, semi-supervised ones are particularly appealing, since they require only jamming-free data, whereas supervised ones require labeled data for every jamming strategy. \ac{AE}-based schemes learn to reconstruct regular signal patterns and flag deviations as anomalies, either on images of the \ac{IQ} samples, which capture localized interference patterns~\cite{hanegraaf2025_arxiv,sciancalepore2023_iotj}, or, through \acp{LSTM-AE}, on their temporal dependencies~\cite{kouchaki2024_wiopt}. \ac{AWGN} is the most common jamming signal, while deceptive jamming, i.e., the replay of legitimate messages, is studied less often~\cite{hanegraaf2025_arxiv,sciancalepore2023_iotj}. Moreover, most works detect strong jamming, which fully prevents the reception of legitimate messages~\cite{nguyen2023_wisec}, and only a few address weak jamming, which affects the link without destroying it~\cite{hanegraaf2025_arxiv,sciancalepore2023_iotj,Bogdanoski,Garnaev,Venkata,Villain}. Weak-jamming detection matters when countermeasures can still be deployed, e.g., when a mobile device approaches a jammer, and is therefore particularly relevant for satellite links, which serve emergency communications and maritime and air navigation.

However, very few works consider early detection of jamming on satellite links, where the signal arrives weaker and propagation differs. Satellite-oriented detection has focused on \ac{GNSS}, where the very low signal power makes raw \ac{IQ} samples hard to exploit, so detectors rely on system-specific metrics, such as the carrier-to-noise ratio and the automatic gain control level~\cite{axell2015_ion}, or on spectrograms~\cite{zhong2024_tim}. Only~\cite{smailes2024_spacecom} considers jamming of a real \ac{LEO} system, and it analyzes the messages still decoded under jamming, but it studies the effect of jamming on \ac{RFF} rather than its detection, while~\cite{salkield2025_wisec,sathaye2022_acsac} focus on attacks and their implications. 
As summarized in Table~\ref{tab:comparison}, the current literature provides limited insight into the physical-layer effects of jamming that remain observable before communication is disrupted. Specifically, the influence of deployment factors such as receiver mobility, propagation environment, jammer distance, and signal type on LEO jamming detectability is currently unexplored. Finally, existing work offers little understanding of whether conventional signal-quality indicators capture the information most relevant for early detection in LEO satellite-IoT links. As a result, the physical factors governing jamming detectability in satellite communications currently remain unexplored.
\section{Reference Scenario and Adversary Model}
\label{sec:sys_advModel}

\begin{figure}[t]
    \centering\includegraphics[width=\columnwidth]{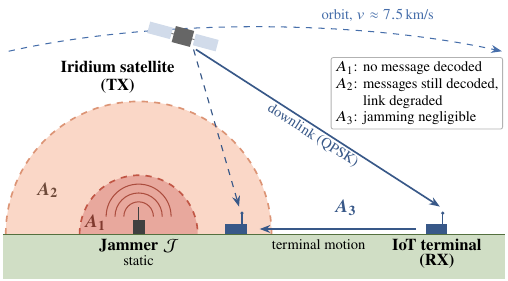}
    \caption{Reference scenario. An Iridium satellite-IoT terminal moves from an unjammed region ($A_3$) to regions subject to increasing jamming power ($A_2$ and $A_1$).}
    \label{fig:systemmodel}
\end{figure}

\textbf{Reference Scenario.} Our reference scenario, depicted in Fig.~\ref{fig:systemmodel}, considers an Iridium satellite-\ac{IoT} deployment. It involves three entities: (i) an Iridium satellite acting as the \ac{TX}, (ii) an \ac{IoT} terminal acting as the \ac{RX} in the considered downlink, and (iii) a ground-based adversary equipped with a \ac{RF} jammer whose objective is to deny the terminal access to the satellite service.

The \ac{RX} terminal is representative of the \ac{M2M} devices served by Iridium~\cite{Iridium_satellite_iot}, i.e., unattended and often battery-powered IoT units installed on shipping containers or remote stations, without terrestrial coverage. Such terminals \emph{are resource-constrained}, with an \ac{SDR}-class front end and an embedded processor with a memory budget of at most a few megabytes. 
Moreover, we consider that no labeled examples of jamming are available for training. We consider terminals that can access the raw \ac{PHY}-layer \ac{IQ} samples of correctly decoded messages, as commercially available \acp{SDR} allow. 
The operating scenario is inherently dynamic. Each Iridium satellite remains visible for only a few minutes, during which the channel geometry evolves, and the carrier undergoes Doppler shifts of several tens of kilohertz, even when the terminal is static. Moreover, a mobile terminal, e.g., on a land vehicle or a vessel, also changes its local propagation environment, which adds amplitude and phase fluctuations to the received signal.
Within this scenario, the jammer partitions the area into the three regions of Fig.~\ref{fig:systemmodel}. In $A_1$, the jamming power prevents the decoding of any message, and the service is unavailable. In $A_2$, the terminal still decodes messages, albeit with degraded quality. In $A_3$, propagation loss attenuates the jammer to a negligible level, and the service is unaffected.

Detection in $A_1$ is trivial and operationally irrelevant, since the absence of decoded messages reveals the attack only after connectivity is lost. The operationally relevant region is therefore $A_2$, where the terminal can still react, but its degradation is ambiguous, since unintentional interference, obstacles, adverse weather, or a low satellite elevation produce similar effects. Unlike obstacles, weather, and a low elevation, which attenuate the signal, a jammer adds interference. This work builds on the hypothesis that the \ac{PHY}-layer samples of the messages still decoded in $A_2$ carry a signature of this interference, whose detection leaves the terminal enough time to trigger a countermeasure, e.g., deferring transmissions, raising an alert over a backup link, or leaving the affected area. Separating a jammer from unintentional broadband interference with the same effect is beyond the scope of this work.

\textbf{Adversary Model.} We consider an adversary $\mathcal{J}$ that aims at \emph{denial of service} against the satellite-\ac{IoT} terminals located in a given area. $\mathcal{J}$ is ground-based and relies on a low-cost \ac{SDR} with an omnidirectional antenna tuned to $f_j = \SI{1622}{\mega\hertz}$, inside the Iridium user band. Once deployed, $\mathcal{J}$ remains static and transmits continuously at the maximum power of its hardware, without knowing the exact location of its targets, as is natural for satellite terminals scattered over wide and sparsely monitored areas.
Under these conditions, $\mathcal{J}$ resorts to constant barrage jamming.
In particular, $\mathcal{J}$ employs one of two waveforms, namely broadband \ac{AWGN} or a \emph{deceptive} waveform that is a replay of previously captured Iridium signals. The latter requires prior knowledge of the link and thus models a more capable adversary.
This constant, non-adaptive jammer matches the reference adversary of the jamming detection literature (Table~\ref{tab:comparison}). Reactive jammers, motivated by tight energy budgets, and overshadowing attackers, which must synchronize with specific transmissions and pursue stealthiness rather than denial of service, are out of scope, as is signal forgery (spoofing), already addressed in the literature~\cite{smailes2024_spacecom,wigchert2025_comnet}.

\textbf{Problem Statement.} Based on the reference scenario and adversary model, we address the timely detection of jamming before communication is disrupted. The detector must rely only on correctly decoded messages, learn legitimate traffic without knowledge of the jamming strategy, remain effective across jamming waveforms and environments, distinguish jamming from the variability induced by propagation and mobility, and fit the computational and memory budget of satellite-\ac{IoT} terminals.

\section{Jamming Detection Strategies}
\label{sec:jem_det_tech}

In this section, we present the considered jamming detection techniques. We consider three \ac{AI}-based detectors used in the context of terrestrial links, i.e., the \ac{IB AE}, the \ac{LSTM-AE}, and an \ac{AE} on raw \ac{IQ} samples, and two statistical baselines adopted in the \ac{GNSS} domain, i.e., a moment-based $C/N_0$ estimator and a symbol-directed \ac{SNR} estimator. All the AI-based detectors treat jamming as an anomaly of the received messages, but they expose it differently, i.e., through spatial structure in the \ac{IB AE}~\cite{sciancalepore2023_iotj,hanegraaf2025_arxiv}, through dependencies along a sequence of samples in the \ac{LSTM-AE}~\cite{kouchaki2024_wiopt}, and through neither in the plain \ac{AE}, which isolates the contribution of both mechanisms. 

Note that the goal of this work is not to identify the best architecture for anomaly detection. Rather, we investigate whether correctly decoded satellite messages contain sufficient information to support timely jamming detection and, if so, which physical-layer characteristics enable such detection.
By comparing detectors that capture local IQ-sample statistics, signal-image representations, and temporal dependencies, we can assess whether the observed detection capabilities depend on a specific architecture or arise from more fundamental traces left by jamming in the received signal. Thus, the considered detectors represent complementary tools for studying jamming detectability in the Iridium LEO satellite-IoT links.
%

\textbf{Image-based Autoencoder.} We adopt the \ac{IB AE} from~\cite{sciancalepore2023_iotj}, which converts blocks of \ac{IQ} samples into images. Given $K$ samples $\{(I_k,Q_k)\}^K_{k=1}$, we divide the constellation plane into a $224 \times 224$ grid and use a histogram $H$ to count how many samples fall in each tile, as in Eq.~\ref{eq:hist}, where $b_I$ and $b_Q$ denote the tile edges along the $I$ and $Q$ axes.
\begin{equation}
    \label{eq:hist}
    H_{i,j}=\sum^K_{k=1}\mathbb{I}(b_{I,i}\leq I_k<b_{I,i+1}) \cdot \mathbb{I}(b_{Q,j}\leq Q_k <b_{Q,j+1}),
\end{equation}
with $\mathbb{I}$ the indicator function. The AE reads bins as pixels of a grey-scale image, whose brightness is the bin count. The grid spans $[-A, A]$ along both axes, with $A$ equal to twice the median, over the training blocks, of the 99th percentile of the pooled values of $|I|$ and $|Q|$ of each block. Samples outside the grid are discarded, the outermost row and column on each side are set to zero, and the histogram is then blurred with a Gaussian kernel with a standard deviation of 2~pixels, truncated at 3 standard deviations, with zero padding. The counts are not divided by $K$, so that, per message, the image also reflects the length of the message. The reference architecture in~\cite{sciancalepore2023_iotj} encodes the image with 8-bit pixels and therefore clips the tile counts at 255. Here, instead, we keep the blurred counts in floating point: the densest tiles, which carry discriminative information, never saturate.

The model is trained on images built from unjammed \ac{IQ} samples by minimizing the \ac{MSE} between the input $x$ and its reconstruction $\hat{x}=g_\phi(f_\theta(x))$. The encoder has $L=16$ logistic-sigmoid neurons, i.e., $z=\sigma(W^{(1)}x+b^{(1)})$, with $W^{(1)} \in \mathbb{R}^{16 \times n}$, $b^{(1)} \in \mathbb{R}^{16}$, $n = 50\,176$, and $\sigma(s)=1/(1+e^{-s})$, while the linear decoder returns $\hat{x}=W^{(2)}z+b^{(2)}$ with $S=50\,176$ outputs. Each pixel $x_i$ is rescaled to $[-1, 1]$ as $\tilde{x}_i = 2(x_i - m_i)/\Delta_i - 1$, where $m_i$ and $\Delta_i$ are its minimum and range over the training set, and pixels that are constant in training ($\Delta_i = 0$) are assigned a unit range. The reconstruction error is computed in this rescaled space, where the network is trained. Since a tile that is rarely occupied in regular traffic has a small range, a jammer that widens the \ac{IQ} cloud and fills it produces a large error. Pixels with a small but non-zero range are not floored, so that they weigh far more than never-occupied ones. The same weighting, however, also amplifies legitimate changes of the cloud, which may explain the less stable operating point of the \ac{IB AE}.

\textbf{LSTM Autoencoder.} The \ac{LSTM-AE} is an encoder-decoder network built on \ac{LSTM} layers~\cite{hochreiter1997}. At each step $t$, the cell state and the hidden state are updated as $c_t = f_t \odot c_{t-1} + i_t \odot g_t$ and $h_t = o_t \odot \tanh(c_t)$, where $\odot$ is the element-wise product, the input, forget, and output gates follow $\star_t = \sigma(W_\star x_t + R_\star h_{t-1} + b_\star)$, with $\star \in \{i, f, o\}$, and the cell candidate follows $g_t = \tanh(W_g x_t + R_g h_{t-1} + b_g)$, with learnable input weights $W$, recurrent weights $R$, and biases $b$. The encoder consists of two \ac{LSTM} layers with 64 and 32 hidden units, which compress the input sequence into a latent representation, while the decoder mirrors it, with 32 and 64 hidden units, and ends with a fully connected layer $\hat{x}_t = W_y h_t + b_y$ that maps the 64 hidden units back to the $I$ and $Q$ values. Each input sequence is a window of 16 consecutive \ac{IQ} samples of a block, with two channels ($I$ and $Q$) and a step of 8 samples between windows, and the score of a block is the mean reconstruction loss of its windows. In both this network and the Raw AE, the windows start at the first sample of the block, so that up to 7 trailing samples are not scored, and a block shorter than 16 samples is extended by repeating it cyclically. Every window lies inside one block, and within a block the samples keep their stored order.

\textbf{Autoencoder on Raw IQ Samples.} To separate the contribution of the AE recurrent structure from that of the image representation, we also consider a dense \ac{AE} operating on raw \ac{IQ} samples~\cite{michelucci2022_arxiv}. This network has the same encoder and decoder structure as the \ac{IB AE} ($L=16$ logistic-sigmoid neurons, linear decoder), but its input is a window of 16 consecutive \ac{IQ} samples ($n = 32$ real inputs, $S = 32$ outputs), with a step of 8 samples between windows. The score of a block is the mean reconstruction error of its windows. Neither the Raw AE nor the \ac{LSTM-AE} normalizes the chunks to unit power, since both operate on samples divided by a fixed constant (Sec.~\ref{sec:evaluation}), and the Raw AE additionally applies the fixed per-feature min--max scaling learned on the training set by the MATLAB autoencoder.

\textbf{Statistical GNSS Baseline Detectors.} Both GNSS baselines operate on \ac{IQ} samples, require no training, and produce a scalar statistic value. Since both statistics decrease under jamming, their score is the negated value.

\subsubsection{Moment-Based \texorpdfstring{$C/N_0$}{C/N0} Estimator}
We use the M2M4 estimator from~\cite{pauluzzi2000,wiesel_cno}, which uses the second and fourth moments of the amplitude of the received IQ samples to estimate $C/N_0$.
Given a block of $K$ complex samples $r[k] = I[k] + jQ[k]$, we compute the second and fourth moments of the received amplitude as $\hat{M}_2 = \frac{1}{K}\sum_{k=1}^{K}|r[k]|^2$ and $\hat{M}_4 = \frac{1}{K}\sum_{k=1}^{K}|r[k]|^4$.
For a constant-modulus signal under \ac{AWGN}, the signal and noise powers follow from the moments as $\hat{S} = \sqrt{2\hat{M}_2^2 - \hat{M}_4}$ and $\hat{N} = \hat{M}_2 - \hat{S}$, yielding the closed-form M2M4 estimate of Eq.~\ref{eq:m2m4_score}:
\begin{equation}
    \Lambda_{\text{C/N}_0}
    = 10\log_{10}\!\left(\frac{\sqrt{2\hat{M}_2^2 - \hat{M}_4}}
                              {\hat{M}_2 - \sqrt{2\hat{M}_2^2 - \hat{M}_4}}\right)
    \;\text{[dB]}.
    \label{eq:m2m4_score}
\end{equation}
Strictly, Eq.~\ref{eq:m2m4_score} estimates the \ac{SNR} per symbol, $\hat{S}/\hat{N}$, which differs from $C/N_0$ by the constant term $10\log_{10}R_s$, with $R_s$ the symbol rate, and therefore ranks the chunks in the same order. Following the GNSS literature, we refer to it as the $C/N_0$ estimator. Moreover, the estimate is real and finite only when $1 < \hat{M}_4/\hat{M}_2^2 < 2$, a condition we use in Sec.~\ref{sec:evaluation} to define the valid chunks.
In regular conditions, the received signal retains the statistics of a \ac{QPSK} constellation and $\Lambda_{\text{C/N}_0}$ tracks the link $C/N_0$. Interference distorts the higher-order statistics, driving the estimate below its unjammed baseline. Thus, a lower estimate indicates a higher probability of jamming.

\subsubsection{Symbol-Directed SNR Estimator}
This method operates at the demodulated symbol level, estimating the noise power from the distance between received samples and their nearest constellation points, without knowledge of the transmitted sequence.
Since the stored samples are already symbol-spaced, no timing recovery is needed. The estimator first removes the unknown channel gain and the residual phase offset due to Doppler and oscillator drift, which would otherwise corrupt the symbol decisions. Both are estimated jointly and without pilots through a single complex coefficient. After normalizing the block to unit power, we alternate hard decisions $\hat{s}^{(i)}[k]$ on the unit-energy \ac{QPSK} constellation $\{(\pm 1 \pm j)/\sqrt{2}\}$, given by the signs of the real and imaginary parts of $r[k]/\hat{\alpha}^{(i-1)}$, with the least-squares update
\begin{equation}
    \hat{\alpha}^{(i)} = \frac{1}{K}\sum_{k} r[k]\,\bigl(\hat{s}^{(i)}[k]\bigr)^{*}
    \label{eq:alpha}
\end{equation}
where $^{*}$ denotes complex conjugation, since $|\hat{s}^{(i)}[k]|^2 = 1$. The first decisions are taken on the normalized block, i.e., with $\hat{\alpha}^{(0)} = 1$, and the update is repeated for $N_{\text{iter}} = 3$ iterations, so that $|\hat{\alpha}|$ captures the gain and $\arg\hat{\alpha}$ the phase. Finally, with $\hat{\alpha} = \hat{\alpha}^{(N_{\text{iter}})}$ and the decisions $\hat{s}[k]$ re-sliced with it, the residual power $\hat{\sigma}_w^2 = \frac{1}{K}\sum_{k} |r[k] - \hat{\alpha}\,\hat{s}[k]|^2$ estimates the noise, and the decision statistic is $\Lambda_{\text{SNR}} = 10\log_{10}(|\hat{\alpha}|^2 / \hat{\sigma}_w^2)$~dB, which, like $\Lambda_{\text{C/N}_0}$, decreases as interference raises the noise power. Per message, both estimators process a single burst, as they assume. In a chunk that spans several messages, however, a single coefficient $\hat{\alpha}$ is fitted over bursts with independent gains and phases, and the amplitude differences between bursts raise the kurtosis $\hat{M}_4/\hat{M}_2^2$, which penalizes both baselines at larger chunk sizes.

\section{Data Collection}
\label{sec:data_collection}
\begin{figure}[t]
    \centering
    \subfloat[]{\includegraphics[width=0.4\columnwidth]{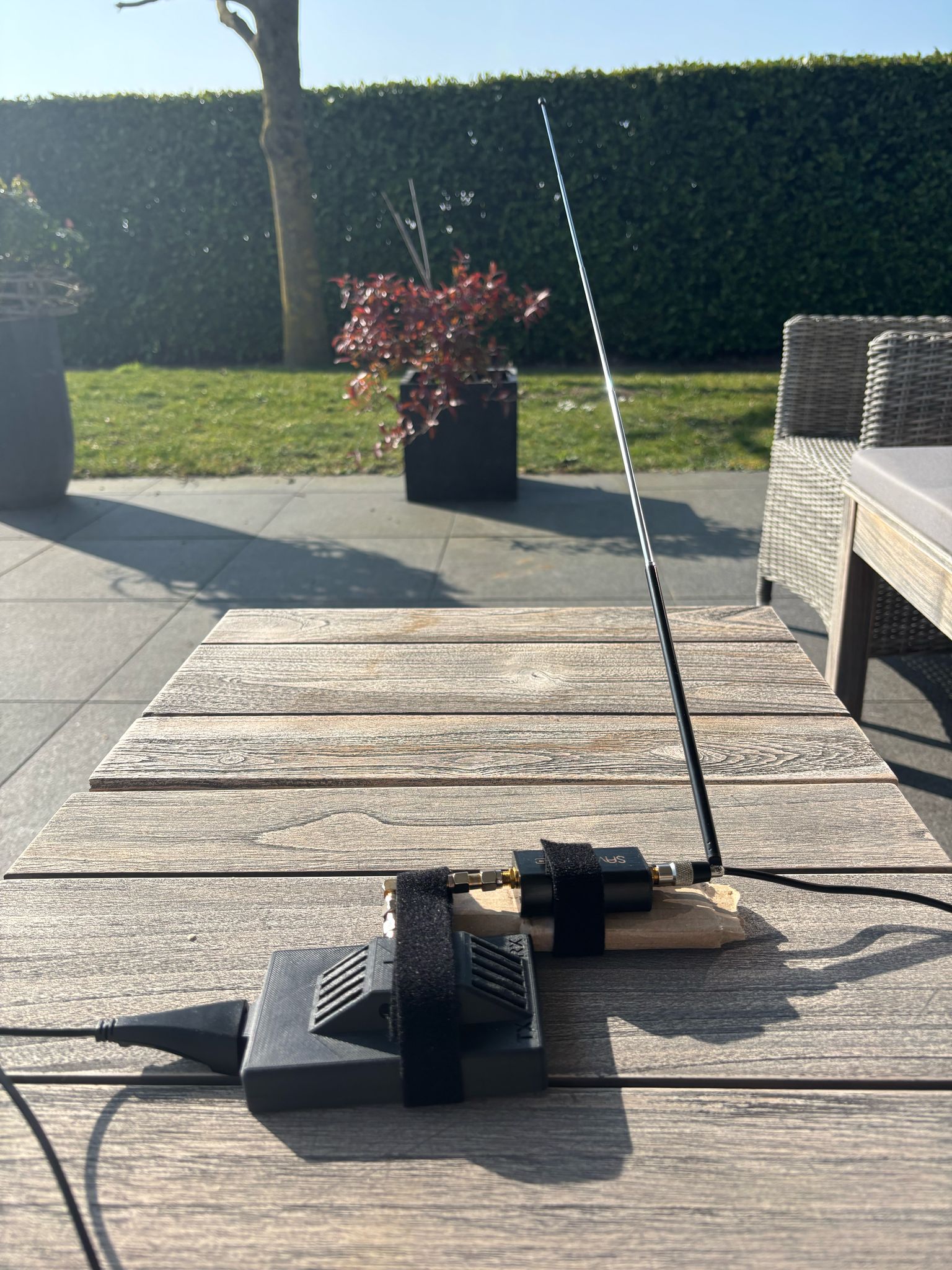}}
    \hfil
    \subfloat[]{\includegraphics[width=0.4\columnwidth]{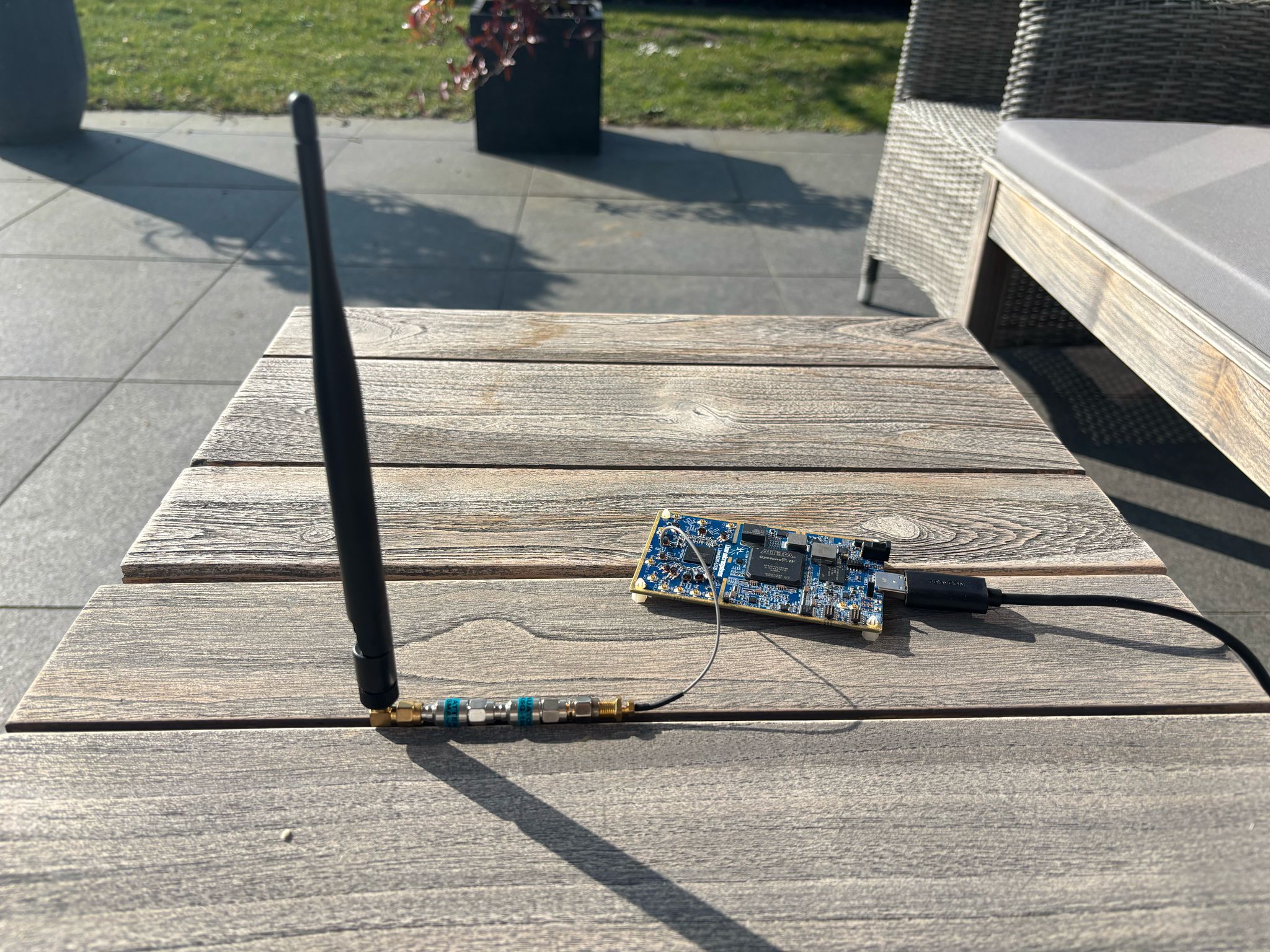}}
    \caption{Experimental setup: Iridium receiver (a) and jammer (b).}
    \label{fig:setup}
\end{figure}

\begin{figure}[t]
    \centering
    \subfloat[]{\includegraphics[width=0.4\columnwidth]{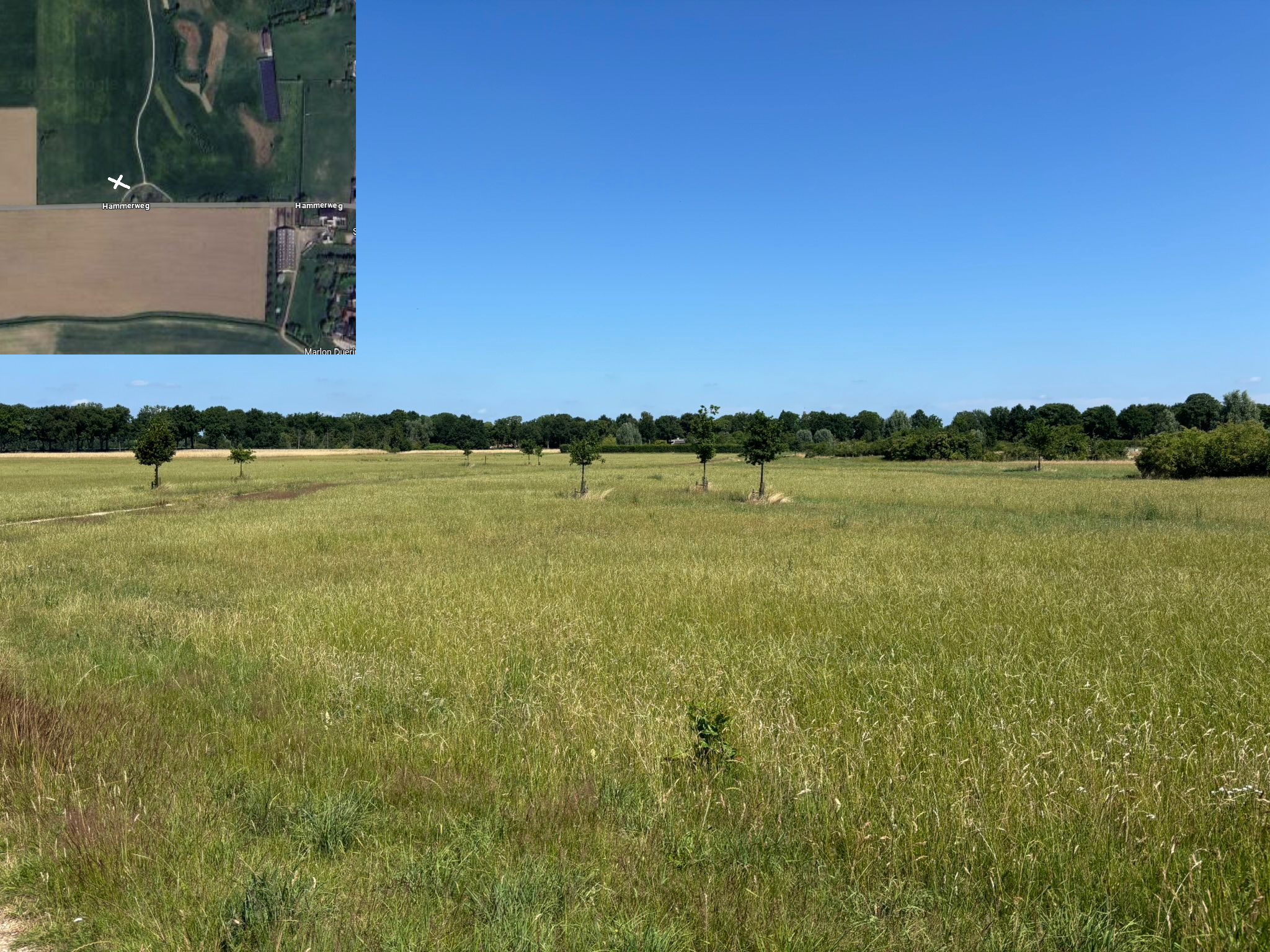}}
    \hfil
    \subfloat[]{\includegraphics[width=0.4\columnwidth]{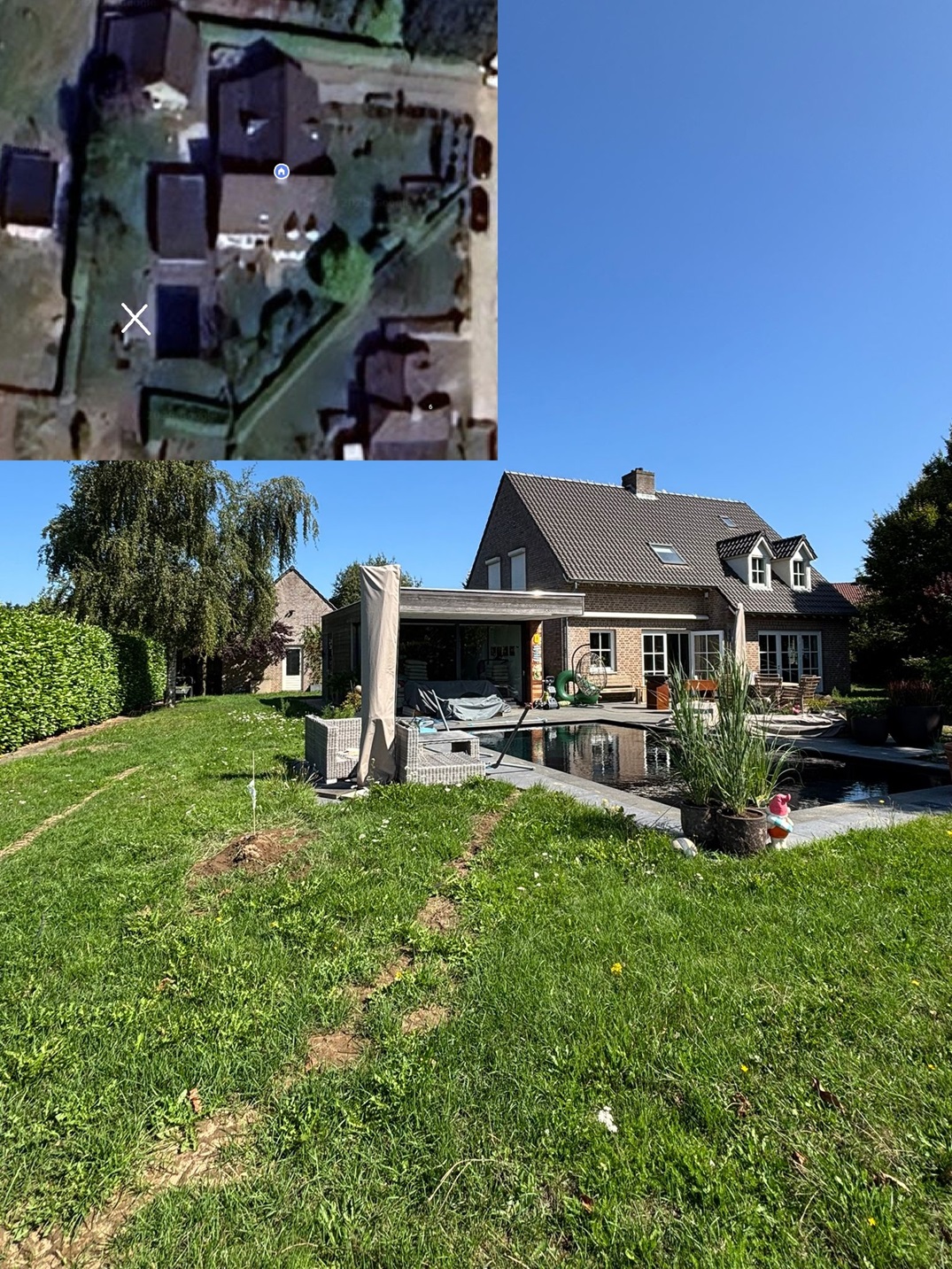}}
    \caption{Measurement environments: (a) low multipath (LM) and (b) high multipath (HM). The X on the satellite image marks the receiver location.}
    \label{fig:envs}
\end{figure}

This section describes the experimental testbed we deployed for data collection, including the hardware and software setups, the measurement scenarios, and the experimental campaigns.

\textbf{Hardware.} Fig.~\ref{fig:setup} shows the experimental setup. The \ac{RX} is a LimeSDR USB~\cite{LimeSDR}, a low-cost \ac{SDR} covering 100~kHz--3.8~GHz with up to 61.44~MHz of bandwidth, preceded by a Nooelec SAWbird+IR \ac{LNA}~\cite{nooelec_sawbird_ir} and connected to an ANT500 antenna. Built entirely from off-the-shelf components, this front end is comparable to that of a low-cost satellite-\ac{IoT} terminal. The jammer is a second LimeSDR USB, with up to 10~dBm of transmit power, followed in all evaluated experiments by a 10~dB attenuator that limits the extent of the jammed area. Both \acp{SDR} are driven by the same HP EliteBook 1040 G4 laptop through USB extension cables of 1.6 to 20~m, while deceptive jamming requires USB 3.0 because of its higher data rate.

\textbf{Software.} On the receiver side, we receive Iridium messages using GNURadio 3.10~\cite{gnuradio} and \emph{gr-Iridium}~\cite{muccc_grIridium_2024}, which we extended to output, alongside each correctly decoded message, the \ac{IQ} samples that precede demodulation. Thus, each stored message carries one \ac{IQ} sample per \ac{QPSK} symbol. For every decoded burst, the receiver log also reports the burst \ac{SNR} and the noise floor estimated on the spectrum, which we later use to measure the \ac{JSR}.
On the jammer side, we implement the two attacks as separate GNURadio flowgraphs. For noise jamming, we use an \emph{Analog Noise Source} of unit complex amplitude, i.e., with a standard deviation per-component of $1/\sqrt{2}$, which feeds a \emph{Multiply Constant} block with a factor of 10, followed by a \emph{LimeSDR Sink}. 
For deceptive jamming, we use a \emph{SigMF Source} which loads a 10-s capture of the Iridium band, repeated in a loop, i.e., a raw IQ replay of the band rather than of isolated bursts, an \emph{IShort To Complex} block which maps its 16-bit samples to $[-1, 1]$, and finally a sink transmitting them. In both cases, the sink transmits at 1622~MHz, so that the 10-MHz replay spans 1617--1627~MHz, i.e., the whole Iridium band except its lowest megahertz. The jammer gain is the digital gain stage of the LimeSDR~\cite{LimeSDR}, set to 73~dB unless stated otherwise. 

\begin{figure}[t]
    \centering
    \includegraphics[width=\columnwidth]{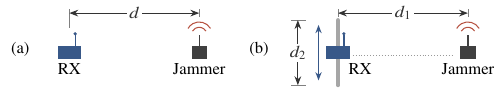}
    \caption{Measurement setup with (a) static and (b) mobile receiver.}
    \label{fig:expsetup}
\end{figure}



\textbf{Measurement and Experiments.} We carry out the measurements in two environments with different multipath conditions, shown in Fig.~\ref{fig:envs}. The first, denoted \emph{low multipath} (LM), is an open field with no major obstruction within 100~m of the \ac{RX}. Apart from ground reflections, multipath fading is minimal, so this environment represents the cleanest channel conditions for a narrowband \ac{LEO} link like Iridium. The second, denoted \emph{high multipath} (HM), is a garden with trees, plants, a pool, and a nearby house. Here, multipath fading is likely, as in the urban-like settings where satellite-\ac{IoT} terminals also operate. We execute tests in both environments where the receiver is either static or mobile, and we refer to each combination as a {\em scenario}, e.g., HM static or LM mobile.
\begin{table}[t]
\centering
\caption{Acquisition experiments.}
\label{tab:experiments}
\setlength{\tabcolsep}{4pt}
\renewcommand{\arraystretch}{1.08}
\begin{adjustbox}{width=\columnwidth}
\begin{tabular}{c l l l l}
\toprule
Exp. & Factor & Scenario & Jammer & Varied parameter (cond.) \\
\midrule
1 & Jammer power & HM static & AWGN & gain 48--73 dB, $d$ = 10 cm (7) \\
\midrule
\multirow{2}{*}{2} & \multirow{2}{*}{Distance} & HM static & AWGN & $d$ = 10--100 cm (10) \\
 & & LM static & AWGN & $d$ = 20--180 cm (9) \\
\midrule
3 & Jamming signal & HM static & Deceptive & $d$ = 20--120 cm (6) \\
\midrule
\multirow{2}{*}{4} & \multirow{2}{*}{RX mobility} & HM mobile & AWGN & $d_1$ = 20--120 cm (6) \\
 & & LM mobile & AWGN & $d_1$ = 20--180 cm (9) \\
\bottomrule
\end{tabular}
\end{adjustbox}
\end{table}
In these scenarios, as summarized in Table~\ref{tab:experiments}, we conduct four acquisition experiments, each varying a single jamming factor:
\begin{itemize}
    \item \textit{Experiment~1 (jammer power).} In the HM static scenario (Fig.~\ref{fig:expsetup}(a)), the jammer is placed at $d = 10$~cm from the \ac{RX}, and its gain is varied over 48, 50, 53, 58, 63, 68, and 73~dB to isolate the effect of its transmit power. 
    \item \textit{Experiment~2 (distance).} In the LM static and HM static scenarios, the noise jammer transmits at 73~dB gain, and its distance $d$ from the \ac{RX} is varied from 20 to 180~cm in 20~cm steps in LM, and from 10 to 100~cm in 10~cm steps in HM. We did not extend the range further for legal aspects, not to affect regular Iridium-IoT users.
    \item \textit{Experiment~3 (jamming signal).} In the HM static scenario, the jammer replays previously captured Iridium signals at $d$ from 20 to 120~cm in 20~cm steps, so that the two jamming signals can be compared at equal distance between 20 and 100~cm.
    \item \textit{Experiment~4 (\ac{RX} mobility).} In the LM mobile and HM mobile scenarios (Fig.~\ref{fig:expsetup}(b)), the \ac{RX} repeatedly traverses a guide rail of length $d_2 = 70$~cm at about 11~m/min, perpendicular to the line joining it to the jammer. The minimum distance $d_1$ between the jammer and the rail is varied from 20 to 180~cm in LM and from 20 to 120~cm in HM, in 20~cm steps.
\end{itemize}

For each measurement session, we collect the correctly decoded Iridium messages, each stored with one \ac{IQ} sample per \ac{QPSK} symbol, together with the receiver log, which reports the burst \ac{SNR} and the noise floor of every decoded burst. 

\section{Evaluation Settings}
\label{sec:evaluation}

\begin{figure*}[!tp]
\centering
\subfloat[HM static]{\includegraphics[width=0.24\textwidth]{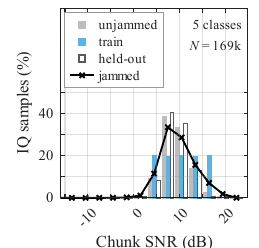}\label{fig:fig00_dataset_snr_a}}
\hfil
\subfloat[LM static]{\includegraphics[width=0.24\textwidth]{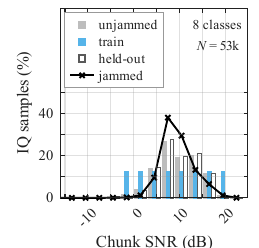}\label{fig:fig00_dataset_snr_b}}
\hfil
\subfloat[HM mobile]{\includegraphics[width=0.24\textwidth]{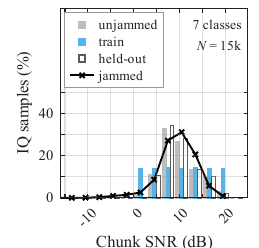}\label{fig:fig00_dataset_snr_c}}
\hfil
\subfloat[LM mobile]{\includegraphics[width=0.24\textwidth]{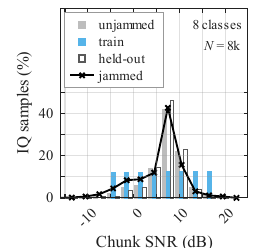}\label{fig:fig00_dataset_snr_d}}
\caption{SNR distribution per message in the four scenarios. Bars: share of IQ samples per 3-dB class in all valid unjammed chunks before the split (unjammed), in the training set (train), and in the held-out test negatives (held-out). Line: jammed AWGN chunks, all distances. $N$: training IQ samples per admitted class.}
\label{fig:dataset}
\end{figure*}

This section describes the dataset and its preprocessing, the detection tests and the considered metrics.


{\bf Data Preprocessing.}
Before being processed by the detectors, the stored traces undergo some preprocessing steps.

We first remove isolated glitch samples. Since legitimate IQ samples rarely exceed 0.3 in absolute value, we discard every sample with $|I| \geq 0.5$ or $|Q| \geq 0.5$. Then, we estimate the \ac{SNR} of each chunk through the M2M4 estimator of Eq.~\ref{eq:m2m4_score}, after excluding samples whose amplitude exceeds three times the chunk median. Chunks are considered valid only if the resulting kurtosis ratio $\hat{M}_4/\hat{M}_2^2$ lies between 1 and 2, as required by the estimator. This operation retains 38,658/38,800 (99.6\%) chunks in HM static, 35,041/37,622 (93.1\%) in LM static, 6,784/7,119 (95.3\%) in HM mobile, and 5,458/10,038 (54.4\%) in LM mobile.

Subsequently, we group the valid chunks into 3-dB SNR classes separately for each scenario, chunk size, and jammer type. Since the unjammed traffic is heavily concentrated in a limited number of SNR classes, directly training on all available unjammed data would allow the detector to learn the typical SNR of the traffic rather than the shape of its \ac{IQ} distribution. To mitigate this effect, we construct one training set per scenario and chunk size by balancing the contribution of the admissible SNR classes extracted from the unjammed data. Each class contributes to the training set the same number of chunks, equal to 80\% of those available in the least populated class. The remaining unjammed chunks are reserved for testing.

Finally, we normalize all \ac{IQ} samples by a common constant, i.e., the median \ac{RMS} amplitude of the training chunks. This approach preserves information related to variations in the received power while keeping the input scale consistent across training and test data.

Fig.~\ref{fig:dataset} shows the resulting message-level \ac{SNR} distributions for the four measurement scenarios. The admitted classes cover 98--99\% of the held-out unjammed samples and 95--100\% of the jammed samples, while the overlap between the jammed and unjammed distributions ranges from 79\% to 93\%. Therefore, \ac{SNR} alone cannot reliably distinguish jammed from unjammed chunks, and any detection capability must arise from other characteristics of the received signal.

\textbf{Jamming Detection Implementation.} We implement all detectors in MATLAB R2024b. 
The Raw AE and the \ac{IB AE} are trained with scaled conjugate gradient for 500 and 250 epochs, respectively, with an L2 regularization of $10^{-4}$ and no sparsity regularization, while the \ac{LSTM-AE} is trained with \texttt{RMSProp} for 100 epochs, with a cyclic learning rate of $10^{-3}$ and mini-batches of 700 windows. For each chunk, we compute a detection score. For the three considered one-class autoencoder-based anomaly detectors, the score is the mean reconstruction error. For the GNSS baselines, the score is the negated value of the estimated statistic, such that larger scores consistently correspond to a higher likelihood of jamming.

{\bf Evaluation Tests.} Using the data collected in Experiments~1--4 (Sec.~\ref{sec:data_collection}), we evaluate the detectors through three complementary tests. In every condition, we treat the valid jammed chunks as positive samples and the held-out unjammed chunks of the same scenario as negative samples. The same negative set is shared across all conditions and detectors evaluated within that scenario.
\begin{itemize}
    \item \textit{Test~1 (chunk size).} Through this test, we investigate the trade-off between detection delay and detection performance. Using the AWGN-jamming conditions of Experiments 1, 2, and 4, we evaluate the detectors on a single decoded message and on chunks containing $c = 250, 500, 1000, 2000$ \ac{IQ} samples. Based on these results, the following analyses focus on two representative operating points: a single message, which minimizes the detection delay, and $c = 1000$, which provides a favorable trade-off between delay and detection performance.

    \item \textit{Test~2 (in-domain detection).} Through this test, we evaluate the detectors in the same scenario used for training. For each condition of Experiments 1-4, we apply the model, normalization constant, and detection threshold derived from the corresponding training set.

    \item \textit{Test~3 (model transferability).} This test evaluates the robustness of the detectors across environments. For both static and mobile receivers, we train the detectors in one environment and test them in the other, i.e., HM-to-LM and LM-to-HM, while retaining the model parameters, normalization constant, and detection threshold of the training environment.
\end{itemize}

{\bf Evaluation Metrics.} We assess the performances of the detectors through two complementary metrics. First, we report the \ac{AUC} of the \ac{ROC} curve, i.e., the probability that a jammed chunk receives a larger score than an unjammed one. Since the \ac{AUC} is independent of the decision threshold and robust to class imbalance~\cite{yang2022_csur}, we use it as our primary measure of separability. We compute the \ac{AUC} through the Mann-Whitney statistic~\cite{mann1947,hanley1982} and estimate a 95\% confidence interval through percentile bootstrap resampling with 200 repetitions, in line with relevant literature~\cite{efron1993}. Moreover, since a deployed jamming detector ultimately operates with a fixed threshold, we also report the \ac{TPR} and \ac{FPR} obtained by placing the threshold at the \ac{P95} of the training scores. We compute the corresponding \ac{FPR} on the held-out unjammed chunks. Therefore, an AUC of 0.5 or lower indicates that the detector cannot detect jamming.

Finally, we characterize the intensity of the jamming observed at the receiver. Neither the jammer transmit power nor the jammer-receiver distance uniquely determines the interference level experienced by the receiver, while the SNR of the stored messages reflects that information only for the bursts that survived jamming. Therefore, we use the receiver noise floor reported by the \ac{SDR} front-end of the receiver, which increases with the received jamming power even when the successfully decoded bursts retain a high SNR. For each measurement session, we compute the median noise floor $\mathrm{NF}$ and compare it with the unjammed reference value $\mathrm{NF}_0$, obtained from the unjammed sessions of the same scenario. From these quantities, we derive the \ac{JSR} and the noise-floor increase $\Delta N$. Throughout the remainder of the paper, we use $\Delta N$ as the primary measure of jamming intensity, because it is defined for all conditions and does not depend on the reference \ac{SNR}.

\begin{figure*}[t]
\centering
\subfloat[HM static (10 dist.)]{\includegraphics[width=0.24\textwidth]{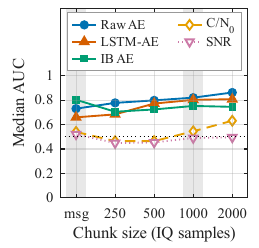}\label{fig:fig01_chunk_awgn_a}}
\hfil
\subfloat[LM static (9 dist.)]{\includegraphics[width=0.24\textwidth]{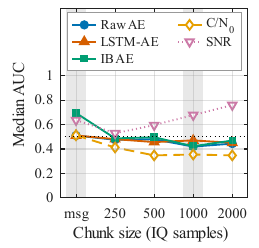}\label{fig:fig01_chunk_awgn_c}}
\hfil
\subfloat[HM mobile (6 dist.)]{\includegraphics[width=0.24\textwidth]{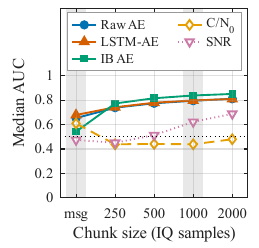}\label{fig:fig01_chunk_awgn_b}}
\hfil
\subfloat[LM mobile (4 dist.)]{\includegraphics[width=0.24\textwidth]{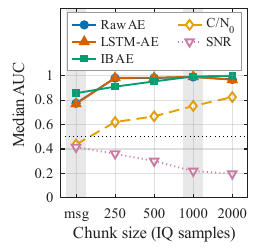}\label{fig:fig01_chunk_awgn_d}}
\caption{Median AUC versus chunk size on AWGN jamming (Test~1), over the distances of each scenario available at all chunk sizes (number in parentheses). Grey bands: chunk sizes used in the rest of the paper.}
\label{fig:chunk}
\end{figure*}

\begin{figure}[t]
\centering
\subfloat[Held-out unjammed chunks]{\includegraphics[width=0.49\columnwidth]{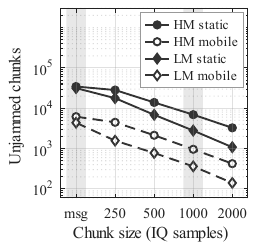}\label{fig:fig01_chunk_awgn_e}}
\hfil
\subfloat[TPR, HM static (10 dist.)]{\includegraphics[width=0.49\columnwidth]{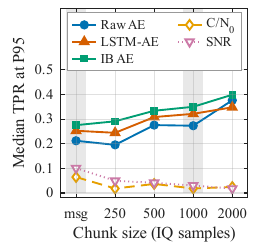}\label{fig:fig01_chunk_awgn_f}}
\caption{Effect of the chunk size on (a) the number of held-out unjammed test chunks per scenario and (b) the median TPR at the P95 threshold, AWGN, HM static.}
\label{fig:chunk_tpr}
\end{figure}

\begin{figure*}[t]
\centering
\subfloat[AUC, per message]{\includegraphics[width=0.32\textwidth]{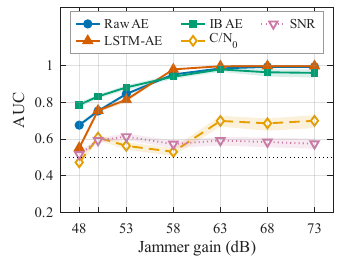}\label{fig:fig03_gain_a}}
\hfil
\subfloat[AUC, $c = 1000$]{\includegraphics[width=0.32\textwidth]{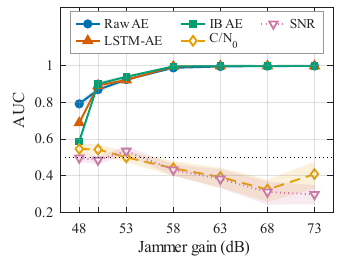}\label{fig:fig03_gain_b}}
\hfil
\subfloat[JSR from the receiver noise floor]{\includegraphics[width=0.32\textwidth]{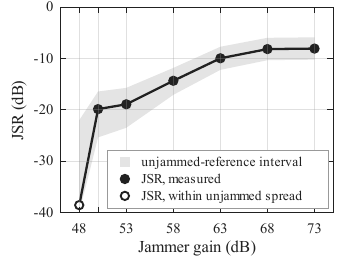}\label{fig:fig03_gain_c}}
\caption{Experiment~1, AWGN jammer gain at $d = 10$~cm, HM static. (a), (b) AUC with 95\% CI. (c) JSR measured from the receiver noise floor, with the interval given by the highest and lowest unjammed session (open marker: within the unjammed spread).}
\label{fig:gain}
\end{figure*}

\begin{figure*}[t]
\centering
\subfloat[HM static, per message]{\includegraphics[width=0.24\textwidth]{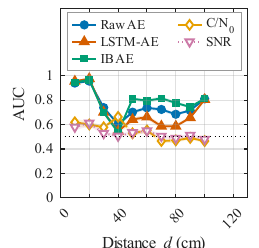}\label{fig:fig02_scenarios_a}}
\hfil
\subfloat[LM static, per message]{\includegraphics[width=0.24\textwidth]{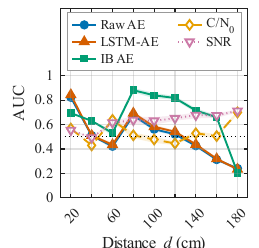}\label{fig:fig02_scenarios_e}}
\hfil
\subfloat[HM mobile, per message]{\includegraphics[width=0.24\textwidth]{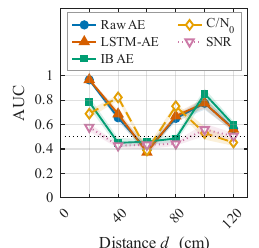}\label{fig:fig02_scenarios_c}}
\hfil
\subfloat[LM mobile, per message]{\includegraphics[width=0.24\textwidth]{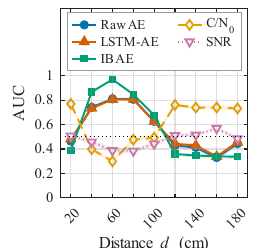}\label{fig:fig02_scenarios_g}}
\\
\subfloat[HM static, $c = 1000$]{\includegraphics[width=0.24\textwidth]{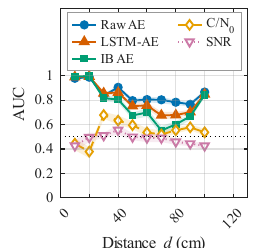}\label{fig:fig02_scenarios_b}}
\hfil
\subfloat[LM static, $c = 1000$]{\includegraphics[width=0.24\textwidth]{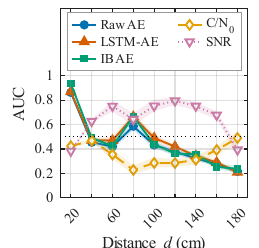}\label{fig:fig02_scenarios_f}}
\hfil
\subfloat[HM mobile, $c = 1000$]{\includegraphics[width=0.24\textwidth]{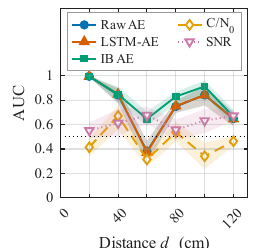}\label{fig:fig02_scenarios_d}}
\hfil
\subfloat[LM mobile, $c = 1000$]{\includegraphics[width=0.24\textwidth]{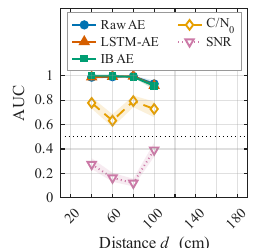}\label{fig:fig02_scenarios_h}}
\caption{Experiments~2 and 4, AUC with 95\% CI of all detectors versus distance under AWGN jamming (static receiver: distance $d$, mobile receiver: minimum distance $d_1$), (a)--(d) per message and (e)--(h) $c = 1000$.}
\label{fig:scenarios}
\end{figure*}

\begin{figure*}[t]
\centering
\subfloat[AWGN, per message]{\includegraphics[width=0.24\textwidth]{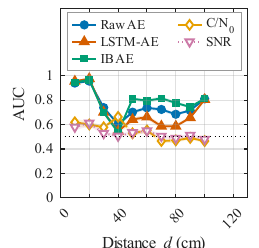}\label{fig:fig06_jammer_type_a}}
\hfil
\subfloat[Deceptive, per message]{\includegraphics[width=0.24\textwidth]{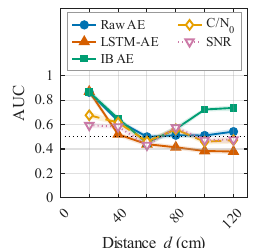}\label{fig:fig06_jammer_type_b}}
\hfil
\subfloat[AWGN, $c = 1000$]{\includegraphics[width=0.24\textwidth]{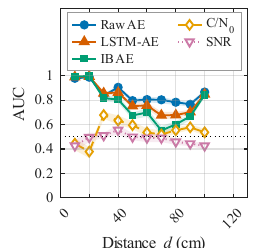}\label{fig:fig06_jammer_type_c}}
\hfil
\subfloat[Deceptive, $c = 1000$]{\includegraphics[width=0.24\textwidth]{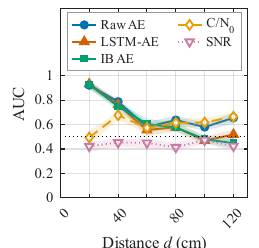}\label{fig:fig06_jammer_type_d}}
\caption{Experiment~3, AUC with 95\% CI of all detectors versus distance under AWGN and deceptive jamming in HM static, (a), (b) per message and (c), (d) $c = 1000$.}
\label{fig:jammertype}
\end{figure*}

\begin{figure}[t]
\centering
\subfloat[per message]{\includegraphics[width=0.49\columnwidth]{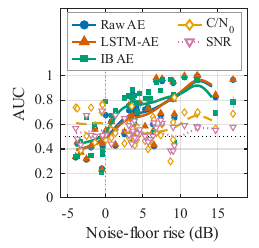}\label{fig:fig04b_auc_vs_dN_a}}
\hfill
\subfloat[$c = 1000$]{\includegraphics[width=0.49\columnwidth]{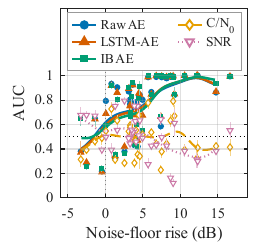}\label{fig:fig04b_auc_vs_dN_b}}
\caption{AUC versus the noise-floor rise $\Delta N$ for all AWGN conditions, (a) per message and (b) $c = 1000$. Markers: one condition with 95\% bootstrap CI (filled: autoencoders, open: GNSS baselines). Lines: Gaussian-kernel trend (1.5-dB bandwidth). Dotted line: $\Delta N = 0$.}
\label{fig:jsr}
\end{figure}

\begin{figure*}[t]
\centering
\subfloat[HM, 20~cm, per message]{\includegraphics[width=0.24\textwidth]{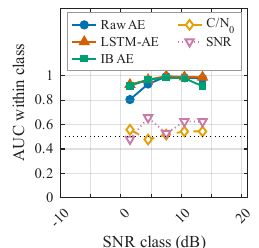}\label{fig:fig05_auc_snr_class_a}}
\hfil
\subfloat[HM, 20~cm, $c = 1000$]{\includegraphics[width=0.24\textwidth]{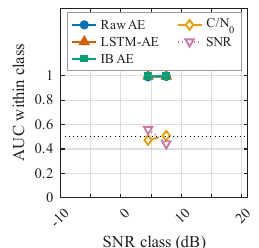}\label{fig:fig05_auc_snr_class_b}}
\hfil
\subfloat[HM, 90~cm, per message]{\includegraphics[width=0.24\textwidth]{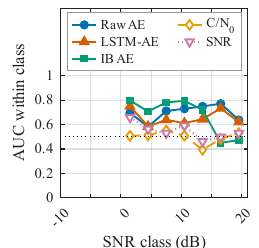}\label{fig:fig05_auc_snr_class_e}}
\hfil
\subfloat[HM, 90~cm, $c = 1000$]{\includegraphics[width=0.24\textwidth]{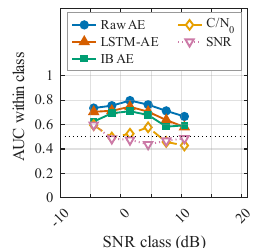}\label{fig:fig05_auc_snr_class_f}}
\\
\subfloat[LM, 20~cm, per message]{\includegraphics[width=0.24\textwidth]{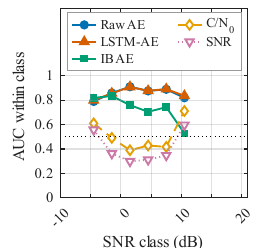}\label{fig:fig05_auc_snr_class_g}}
\hfil
\subfloat[LM, 20~cm, $c = 1000$]{\includegraphics[width=0.24\textwidth]{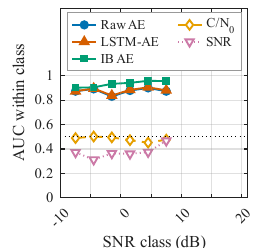}\label{fig:fig05_auc_snr_class_h}}
\hfil
\subfloat[LM, 120~cm, per message]{\includegraphics[width=0.24\textwidth]{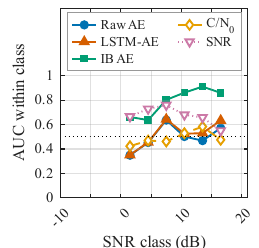}\label{fig:fig05_auc_snr_class_i}}
\hfil
\subfloat[LM, 120~cm, $c = 1000$]{\includegraphics[width=0.24\textwidth]{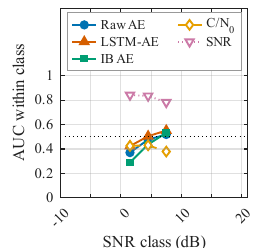}\label{fig:fig05_auc_snr_class_j}}
\caption{AUC within each 3-dB SNR class, computed between jammed and unjammed chunks of the same class, AWGN, static receiver, HM (top) and LM (bottom).}
\label{fig:snrclass}
\end{figure*}

\section{Results and Analysis}
\label{sec:exp_results} 
We first discuss in Sec.~\ref{sec:exp_timeliness} the effect of the chunk size (Test~1), then in Sec.~\ref{sec:exp_detect} the in-domain detection (Test~2). In Sec.~\ref{sec:exp_phy}, we provide evidence on the physical factors governing detectability and why signal-quality metrics fail, and finally, in Sec.~\ref{sec:exp_deploy}, we analyze the operationalization of the models (Test~3) and their cost.

\subsection{Detection Timeliness vs. Available Information}
\label{sec:exp_timeliness}

Fig.~\ref{fig:chunk} reports the median \ac{AUC} of all considered detectors with various chunk sizes in all considered scenarios.
From the reported results, we note that larger chunks generally improve the detection capability of all autoencoder-based physical-layer detectors. For example, the median AUC of the Raw AE increases from approximately 0.68 when operating on individual messages to about 0.80 at $c=1000$, while the LSTM-AE reaches almost its maximum performance already at the same chunk size.
This improvement occurs because longer observation windows expose a larger fraction of the distortions introduced by jamming, allowing the detectors to characterize the received signal more accurately. 
However, as shown in Fig.~\ref{fig:chunk_tpr}, enlarging the chunk size also reduces the number of unjammed test chunks available for analysis. Indeed, larger chunks require larger observation interval before a decision can be made, thereby reducing timeliness, and this occurs with marginal performance gains. 
For example, the LSTM-AE reaches 99\% of its final gain already at $c=1000$.
Therefore, in the following, we consider only two reference chunk sizes, i.e., a single received message, which minimizes the detection delay, and $c = 1000$, as a compromise between the number of test chunks and the detection accuracy.

\subsection{In-Domain Jamming Detectability}
\label{sec:exp_detect}

Fig.~\ref{fig:gain} reports the detection performance of all considered detectors obtained in Experiment 1 as a function of the jammer gain, at $d = 10$~cm in HM static. Subfigures (a) and (b) report the AUC for a single message and for $c=1000$, respectively, while subfigure (c) reports the JSR estimated from the receiver noise floor. 
We note that the performances of the three one-class autoencoder-based anomaly detectors increase almost monotonically with the jammer gain. From 58 dB onwards, all three detectors achieve an AUC of at least 0.94, whereas the GNSS baselines remain substantially lower. At the highest jammer gains and $c=1000$, the two baselines even fall below random guessing. This behavior occurs because increasing the jammer power primarily increases the interference observed at the receiver, which directly affects the physical-layer signal processed by the autoencoders. Conversely, the signal-quality metrics used by the GNSS baselines remain largely unaffected. The key evidence appears in Fig.~\ref{fig:gain}(c): while the measured JSR increases from approximately -20 dB to -8 dB, the SNR of the decoded bursts remains close to 20 dB throughout the experiment. Therefore, the jammer becomes visible through the receiver noise floor, rather than through the SNR of the received messages.

Fig.~\ref{fig:scenarios} extends the analysis to distance and receiver mobility, reporting the effect of jammer distance in Experiments 2 and 4 for static and mobile receivers.
We observe that the autoencoders outperform the GNSS baselines in most tested conditions. The advantage is particularly clear in HM static ((a), (e)) and HM mobile ((c), (g)), where the best autoencoder achieves a higher AUC than the best GNSS baseline in 98\% and 80\% of the condition-chunk pairs, respectively. A second observation is that the curves are highly irregular. Detection performance does not decrease monotonically with the jammer distance, and in some cases a larger distance yields a higher AUC than a shorter one. This behavior indicates that distance alone does not determine detectability. Receiver mobility does not fundamentally hinder detection either. For example, the AUC achieved by the Raw AE and the LSTM-AE remains almost unchanged between HM static and HM mobile at comparable distances. Therefore, neither distance nor mobility directly governs detectability, and their effect must be mediated by another quantity observed at the receiver.


To investigate this further, Fig.~\ref{fig:jammertype} isolates the effect of the jamming signal, comparing \ac{AWGN} and deceptive jamming in HM static (Experiment~3) at equal distances. 
We notice that deceptive jamming is generally harder to detect than AWGN. Across the five chunk sizes and the distances common to both experiments, the deceptive waveform yields a lower AUC in 72 out of 75 comparisons. This behavior occurs because the replayed waveforms raise the receiver noise floor much less effectively than AWGN. The corresponding receiver logs show that a measurable noise-floor increase is observed only at 20 cm, while from 60 cm to 120 cm the effect becomes negligible. However, the results in Fig.~\ref{fig:jammertype} also suggest that waveform characteristics play a role beyond the received interference power. Even at 20 cm, where deceptive jamming produces a JSR approximately 2 dB larger than that of AWGN, the autoencoders still achieve a lower AUC. Therefore, the received jamming power is the dominant factor governing jamming detectability, but the specific waveform can influence detection performance as well.

\subsection{Physical Factors Governing Detectability}
\label{sec:exp_phy}

The previous experiments suggest that detection performance cannot be explained solely by jammer power, distance, waveform, or mobility. Therefore, we investigate whether a common physical quantity can explain the seemingly diverse behaviors observed across all experiments. To this aim, Fig.~\ref{fig:jsr} aggregates all AWGN-jamming conditions from Experiments 1, 2, and 4 and reports the AUC as a function of the noise-floor increase $\Delta N$ (Sec.~\ref{sec:data_collection}).
We notice that conditions originating from different experiments, environments, receiver mobility patterns, distances, and jammer powers follow a common trend when expressed through $\Delta N$. In particular, with a chunk size $c=1000$, all three autoencoder-based anomaly detectors achieve an AUC between 0.84 and 1.00 whenever the noise floor increases by at least 9 dB. Conversely, when the jammer fails to raise the noise floor above the unjammed reference level ($\Delta N \le 0$), no autoencoder achieves an AUC above 0.66. This indicates that variables such as jamming distance, transmit power, and receiver mobility influence detectability only indirectly. What ultimately determines the visibility of the attack is the amount of interference reaching the receiver front-end, which manifests as an increase of the measured noise floor.
Therefore, the increase of the receiver noise floor emerges as the primary physical factor governing jamming detectability in the analyzed Iridium satellite-IoT link. Power, distance, and mobility all affect detection only through their influence on this quantity.

Fig.~\ref{fig:snrclass} further reports the AUC achieved within individual 3-dB SNR classes. In each class, jammed chunks are compared only against unjammed chunks belonging to the same SNR interval.
From the results, we observe that the autoencoders remain capable of separating jammed and unjammed satellite traffic even after controlling for the SNR. For example, at 20 cm in HM, they achieve an AUC between 0.81 and 1.00 per message and between 0.99 and 1.00 at $c=1000$. Conversely, the GNSS baselines remain close to random guessing in most classes.
These results indicate that jamming introduces detectable distortions beyond conventional signal-quality metrics. At the same time, the substantial overlap between the jammed and unjammed SNR distributions shown earlier in Fig.~\ref{fig:dataset} suggests that detecting jamming by only observing successfully decoded signals biases the observed signal-quality measurements toward values typical of regular traffic. Consequently, the decoded messages retain SNR characteristics similar to those observed without jamming, while the receiver noise floor and the IQ distribution reveal more effectively the presence of interference. Although we did not log IQ samples of undecoded bursts, this interpretation is also consistent with the results reported in Figs.~\ref{fig:gain}(c), and~\ref{fig:jsr}.


\subsection{Deployment Implications}
\label{sec:exp_deploy}

While the previous analysis assessed detection capability through the AUC, deployment ultimately requires selecting a fixed operating point. To this aim, Table~\ref{tab:auc_subset} summarizes the in-domain \ac{AUC} on a subset of reference experimental conditions. 
The results show that the detection performance generally follows the trends observed in the AUC analysis. Conditions associated with a large increase in the receiver noise floor typically achieve high detection rates, whereas jamming producing a limited impact on the receiver remains difficult to detect.
However, the relationship between AUC and TPR is not always straightforward. Among the conditions of Table~\ref{tab:auc_subset} where $\Delta N \geq 9$~dB, at $c = 1000$ the Raw AE and LSTM-AE achieve a TPR of 0.98--1.00 at $g = 73$~dB and in HM mobile at $d_1 = 20$~cm (AUC 0.99--1.00), but their TPR falls to 0.63--0.84 in HM static at $d = 10$~cm (AUC 0.98--0.99) and to 0.38--0.42 in LM static at $d = 20$~cm (AUC 0.86--0.87). By contrast, the IB AE achieves a TPR between 0.90 and 1.00 in all four conditions (AUC 0.93--1.00).
%
This behavior suggests that the P95 threshold may be conservative for the Raw AE and LSTM-AE. Since the training procedure balances the SNR classes, the training set over-represents the least populated classes, which also tend to yield larger anomaly scores. Consequently, the resulting threshold may be higher than the one that would be obtained from naturally distributed unjammed traffic. This interpretation is supported by the in-domain FPR values reported in Table~\ref{tab:transfer} (see below), whose medians remain below the intended 0.05 level for both the Raw AE (0.02) and the LSTM-AE (0.03).
Therefore, deployment performance depends not only on the intrinsic separability of jammed and unjammed traffic, but also on how the operating threshold is calibrated from regular traffic.

\begin{table}[!t]
\centering
\caption{In-domain AUC (Test~2) on a subset of conditions, per message and $c = 1000$. Bold: best detector per row and chunk size.}
\label{tab:auc_subset}
\begin{adjustbox}{width=\columnwidth}
\begin{tabular}{l ccccc c ccccc}
\toprule
 & \multicolumn{5}{c}{Per message} & & \multicolumn{5}{c}{$c = 1000$} \\
\cmidrule(lr){2-6}\cmidrule(lr){8-12}
Condition & Raw & LSTM & IB & $C/N_0$ & SNR & & Raw & LSTM & IB & $C/N_0$ & SNR \\
\midrule
\multicolumn{12}{l}{\textit{Exp.~1: HM static, AWGN, $d = 10$~cm}}\\
$g=48$\,dB & 0.68 & 0.55 & \textbf{0.79} & 0.47 & 0.52 &  & \textbf{0.79} & 0.69 & 0.59 & 0.55 & 0.50 \\
$g=58$\,dB & 0.95 & \textbf{0.98} & 0.94 & 0.53 & 0.57 &  & 0.99 & \textbf{1.00} & \textbf{1.00} & 0.44 & 0.43 \\
$g=73$\,dB & \textbf{1.00} & \textbf{1.00} & 0.96 & 0.70 & 0.58 &  & \textbf{1.00} & \textbf{1.00} & \textbf{1.00} & 0.41 & 0.30 \\
\midrule
\multicolumn{12}{l}{\textit{Exp.~2: HM static, AWGN}}\\
$d=10$\,cm & 0.94 & \textbf{0.95} & 0.94 & 0.62 & 0.58 &  & 0.98 & \textbf{0.99} & \textbf{0.99} & 0.44 & 0.43 \\
$d=50$\,cm & 0.70 & 0.64 & \textbf{0.81} & 0.53 & 0.53 &  & \textbf{0.80} & 0.75 & 0.67 & 0.59 & 0.50 \\
$d=100$\,cm & \textbf{0.81} & 0.80 & \textbf{0.81} & 0.46 & 0.48 &  & \textbf{0.87} & 0.85 & 0.84 & 0.54 & 0.43 \\
\multicolumn{12}{l}{\textit{Exp.~2: LM static, AWGN}}\\
$d=20$\,cm & 0.82 & \textbf{0.84} & 0.69 & 0.57 & 0.55 &  & 0.86 & 0.87 & \textbf{0.93} & 0.42 & 0.38 \\
$d=100$\,cm & 0.56 & 0.58 & \textbf{0.84} & 0.48 & 0.63 &  & 0.43 & 0.49 & 0.44 & 0.29 & \textbf{0.75} \\
$d=180$\,cm & 0.24 & 0.23 & 0.21 & 0.70 & \textbf{0.71} &  & 0.22 & 0.21 & 0.24 & \textbf{0.49} & 0.39 \\
\midrule
\multicolumn{12}{l}{\textit{Exp.~3: HM static, deceptive}}\\
$d=20$\,cm & 0.86 & \textbf{0.87} & \textbf{0.87} & 0.68 & 0.59 &  & 0.92 & \textbf{0.93} & \textbf{0.93} & 0.49 & 0.42 \\
$d=60$\,cm & \textbf{0.50} & 0.44 & 0.45 & 0.46 & 0.43 &  & 0.58 & 0.55 & \textbf{0.60} & 0.57 & 0.45 \\
$d=120$\,cm & 0.54 & 0.38 & \textbf{0.74} & 0.47 & 0.48 &  & 0.65 & 0.52 & 0.45 & \textbf{0.66} & 0.42 \\
\midrule
\multicolumn{12}{l}{\textit{Exp.~4: HM mobile, AWGN}}\\
$d_1=20$\,cm & \textbf{0.97} & \textbf{0.97} & 0.78 & 0.69 & 0.58 &  & \textbf{0.99} & \textbf{0.99} & \textbf{0.99} & 0.41 & 0.55 \\
$d_1=60$\,cm & 0.39 & 0.37 & \textbf{0.46} & 0.39 & 0.43 &  & 0.38 & 0.37 & 0.65 & 0.31 & \textbf{0.68} \\
$d_1=120$\,cm & 0.57 & 0.56 & \textbf{0.60} & 0.45 & 0.51 &  & 0.65 & 0.64 & 0.66 & 0.46 & \textbf{0.67} \\
\multicolumn{12}{l}{\textit{Exp.~4: LM mobile, AWGN}}\\
$d_1=40$\,cm & 0.74 & 0.73 & \textbf{0.87} & 0.40 & 0.46 &  & 0.99 & 0.99 & \textbf{1.00} & 0.78 & 0.27 \\
$d_1=100$\,cm & 0.63 & 0.62 & \textbf{0.68} & 0.49 & 0.44 &  & \textbf{0.93} & 0.92 & 0.92 & 0.73 & 0.39 \\
\bottomrule
\end{tabular}
\end{adjustbox}
\end{table}

Table~\ref{tab:transfer} particularly reports the transferability results obtained when the investigated detectors are trained in one environment and evaluated in another. We observe that the Raw AE and the LSTM-AE largely preserve their ranking capability after transfer. Across a total of 155 paired tests, the median $|\Delta\mathrm{AUC}|$ is only 0.007 for the Raw AE and 0.029 for the LSTM-AE, with Spearman correlations between in-domain and transfer AUC values of 1.00 and 0.97, respectively. In contrast, the IB AE exhibits a significantly larger variation ($|\Delta\mathrm{AUC}|$ = 0.095) and a lower correlation (0.55).
This result indicates that the physical-layer artifacts of jamming captured by the Raw AE and the LSTM-AE remain largely consistent across propagation environments. However, the same robustness does not extend to the operating threshold. When transferring from HM static to LM static, the FPR of the Raw AE increases from 0.04 to 0.38 per message and from 0.04 to 0.61 at $c=1000$. Similar increases are observed for the LSTM-AE, whose FPR reaches 0.47 and 0.68, respectively.
This asymmetry occurs because the decision taken by the detector depends primarily on the distortions introduced by jamming, whereas threshold calibration depends on the statistical properties of unjammed traffic. As a result, the detector can often preserve its ability to separate jammed and unjammed observations even when the threshold learned in the source environment becomes inappropriate in the target one.
Therefore, detector transfer appears considerably easier than threshold transfer. In practice, the trained model can often be reused across environments, whereas the operating threshold should be recalibrated using unjammed traffic collected at the deployment site.


\begin{table}[!t]
\centering
\caption{Model transferability (Test~3): median $|\Delta\mathrm{AUC}|$ and FPR at the P95 threshold (target 0.05) of the in-domain model and of the model trained in the other environment, on the same test data.}
\label{tab:transfer}
\begin{adjustbox}{width=\columnwidth}
\begin{tabular}{l l cc cc cc}
\toprule
 & & \multicolumn{2}{c}{$|\Delta\mathrm{AUC}|$} & \multicolumn{2}{c}{FPR in-domain} & \multicolumn{2}{c}{FPR transfer} \\
\cmidrule(lr){3-4}\cmidrule(lr){5-6}\cmidrule(lr){7-8}
Test data (model) & Detector & msg. & 1000 & msg. & 1000 & msg. & 1000 \\
\midrule
\multirow{3}{*}{\shortstack[l]{HM static\\(LM static)}} & Raw AE & 0.003 & 0.030 & 0.01 & 0.02 & 0.00 & 0.00 \\
 & LSTM-AE & 0.057 & 0.042 & 0.02 & 0.02 & 0.00 & 0.00 \\
 & IB AE & 0.015 & 0.094 & 0.02 & 0.05 & 0.03 & 0.00 \\
\midrule
\multirow{3}{*}{\shortstack[l]{LM static\\(HM static)}} & Raw AE & 0.036 & 0.014 & 0.04 & 0.04 & 0.38 & 0.61 \\
 & LSTM-AE & 0.059 & 0.048 & 0.04 & 0.06 & 0.47 & 0.68 \\
 & IB AE & 0.031 & 0.109 & 0.09 & 0.14 & 0.18 & 0.62 \\
\midrule
\multirow{3}{*}{\shortstack[l]{HM mobile\\(LM mobile)}} & Raw AE & 0.009 & 0.004 & 0.03 & 0.03 & 0.00 & 0.01 \\
 & LSTM-AE & 0.022 & 0.024 & 0.03 & 0.03 & 0.00 & 0.00 \\
 & IB AE & 0.137 & 0.284 & 0.08 & 0.16 & 0.10 & 0.03 \\
\midrule
\multirow{3}{*}{\shortstack[l]{LM mobile\\(HM mobile)}} & Raw AE & 0.025 & 0.002 & 0.06 & 0.02 & 0.33 & 0.13 \\
 & LSTM-AE & 0.020 & 0.003 & 0.06 & 0.01 & 0.33 & 0.10 \\
 & IB AE & 0.054 & 0.517 & 0.15 & 0.23 & 0.47 & 0.99 \\
\bottomrule
\end{tabular}
\end{adjustbox}
\end{table}

Finally, we investigate the resource requirements of the various investigated detectors, as reported in Table~\ref{tab:cost}.
The Raw AE contains only 1,072 trainable parameters, compared with 62,850 for the LSTM-AE and 1,655,824 for the IB AE. Furthermore, the Raw AE operates directly on IQ samples through two lightweight matrix-vector multiplications, whereas the IB AE additionally requires constructing a $ 224\times224$ image representation.
These large complexity differences do not translate into equally large differences in performance. Across both the in-domain and transferability evaluations, the Raw AE consistently achieves results comparable to those of the LSTM-AE despite a computational footprint that is more than an order of magnitude smaller.
This result suggests that the information required for jamming detection is contained primarily in local distortions of the IQ distribution, which can already be captured by a lightweight feed-forward architecture. More sophisticated temporal models provide limited additional benefit in the considered scenarios.

Overall, the Raw AE offers the most favorable trade-off between detection capability, transferability, and implementation cost, making it the most suitable candidate for deployment on resource-constrained satellite-IoT terminals.

\begin{table}[!t]
\centering
\caption{Resource requirements of the autoencoders. Memory in 32-bit floating point, including the input scaling. Operations: multiply--accumulate operations per detection decision, excluding activations.}
\label{tab:cost}
\begin{adjustbox}{width=\columnwidth}
\begin{tabular}{l r r r r}
\toprule
 & Trainable & Memory & \multicolumn{2}{c}{Operations per decision} \\
\cmidrule(lr){4-5}
Detector & parameters & & Per message & $c = 1000$ \\
\midrule
Raw AE  & 1\,072      & 4.5~kB & 26~k  & 127~k \\
LSTM-AE & 62\,850     & 251~kB & 25~M  & 123~M \\
IB AE   & 1\,655\,824 & 7.0~MB & 2.9~M & 2.9~M \\
\bottomrule
\end{tabular}
\end{adjustbox}
\end{table}

\section{Conclusion}
\label{sec:conclusion}

This paper investigated to what extent jamming affecting Iridium satellite-IoT links can be detected timely, analyzing the physical layer of correctly decoded messages, enabling detection before communication is disrupted.
Overall, our results provide three main insights. First, jamming detectability is governed primarily by the increase of the receiver noise floor rather than by individual attack parameters such as jammer power, distance, or receiver mobility. Second, conventional GNSS-inspired indicators based on SNR and C/N$_0$ provide limited detection capability because correctly decoded messages often retain favorable signal-quality measurements despite ongoing interference. Third, anomaly detectors operating directly on IQ samples remain effective even within a single SNR class, showing that jamming leaves characteristic physical-layer distortions that are not captured by traditional signal-quality metrics.
These findings demonstrate that early jamming detection in satellite-IoT systems is feasible already from chunks of 1000 samples (i.e., 40~ms of received signal at 25{,}000 samples per second), but requires exploiting physical-layer information rather than conventional signal-quality indicators. Among the evaluated approaches, the Raw AE offers the most favorable trade-off between detection capability, transferability, and implementation cost.

More broadly, this work opens new directions for research on jamming robustness in satellite communications. Future work should investigate whether the observed relationship between receiver noise-floor increase and detectability extends to other satellite systems, receivers, and interference strategies, as well as how physical-layer distortions can be leveraged to design more explainable and transferable detection mechanisms.

\bibliographystyle{IEEEtran}
\balance
\bibliography{bibliography}

\end{document}

%% file: acronyms.tex
\acrodef{AI}{Artificial Intelligence}
\acrodef{RF}{Radio Frequency}
\acrodef{RFF}{Radio Frequency Fingerprinting}
\acrodef{RFFI}{Radio Frequency Fingerprint Identification}
\acrodef{PHY}{Physical}
\acrodef{BPSK}{Binary Phase Shift Keying}
\acrodef{SDR}{Software Defined Radio}
\acrodef{CNN}{Convolutional Neural Network}
\acrodef{DL}{Deep Learning}
\acrodef{HPC}{High-Performance Computing}
\acrodef{ML}{Machine Learning}
\acrodef{M2M}{Machine-to-Machine}
\acrodef{AML}{Adversarial Machine Learning}
\acrodef{PLC}{Power Line Communication}
\acrodef{FGSM}{Fast Gradient Signed Method}
\acrodef{GAN}{Generative Adversarial Networks}
\acrodef{IoT}{Internet of Things}
\acrodef{AGC}{Automatic Gain Control}
\acrodef{BER}{Bit Error Rate}
\acrodef{GPP}{General-Purpose Processor}
\acrodef{DSP}{Digital Signal Processing}
\acrodef{QPSK}{Quadrature Phase Shift Keying}
\acrodef{RRC}{Root Raised Cosine}
\acrodef{IQ}{In-Phase Quadrature}
\acrodef{RGB}{Red-Green-Blue}
\acrodef{UAV}{Unmanned Aerial Vehicles}
\acrodef{XAI}{eXplainable AI}
\acrodef{IoU}{Intersection over Union}
\acrodef{Grad-CAM}{Gradient-weighted Class Activation Mapping}
\acrodef{LIME}{Local Interpretable Model-agnostic Explanations}
\acrodef{SHAP}{SHapley Additive exPlanations}
\acrodef{GCS}{Ground Control Station}
\acrodef{ReLU}{Rectified Linear Unit}
\acrodef{CSI}{Channel State Information}
\acrodef{LS}{Least-Squares}
\acrodef{MMSE}{Minimum Mean-Square Error}
\acrodef{MSE}{Mean Squared Error}
\acrodef{OFDM}{Orthogonal Frequency-Division Multiplexing}
\acrodef{PCA}{Principal Component Analysis}
\acrodef{RSSI}{Received Signal Strength Indicator}
\acrodef{SVM}{Support Vector Machine}
\acrodef{ISAC}{Integrated Sensing and Communications}
\acrodef{SSB}{Synchronization Signal Blocks}
\acrodef{PSK}{Phase-Shift Keying}
\acrodef{ROC}{Receiver Operating Characteristic}
\acrodef{AUC}{Area Under the Curve}
\acrodef{FPR}{False Positive Rate}
\acrodef{DQPSK}{Differential Quadrature Phase Shift Keying}
\acrodef{LEO}{Low Earth Orbit}
\acrodef{AE}{Autoencoder}
\acrodef{LSTM}{Long Short-Term Memory}
\acrodef{RNN}{Recurrent Neural Network}
\acrodef{RX}{Receiver}
\acrodef{TX}{Transmitter}
\acrodef{MRQ}{Main Research Question}
\acrodef{RQ}{Research Question}
\acrodef{OCC}{One-Class Classification}
\acrodef{AWGN}{Additive White Gaussian Noise}
\acrodef{SNR}{Signal-to-Noise Ratio}
\acrodef{LSTM-AE}{Long Short-Term Memory Autoencoder}
\acrodef{FP}{False Positive}
\acrodef{TP}{True Positive}
\acrodef{TPR}{True Positive Rate}
\acrodef{FN}{False Negative}
\acrodef{TN}{True Negative}
\acrodef{biLSTM}{bidirectional LSTM}
\acrodef{LNA}{Low Noise Amplifier}
\acrodef{IB AE}{Image-Based Autoencoder}
\acrodef{CAE}{Convolutional Autoencoder}
\acrodef{IRA}{Iridium Ring Alert}
\acrodef{GNSS}{Global Navigation Satellite Systems}